# Polyrhythmic Bimanual Coordination Training using Haptic Force Feedback

Ramy A. Mounir, and Kyle B. Reed, *Member, IEEE*

*Abstract*— It is challenging to develop two thoughts at the same time or perform two uncorrelated motions simultaneously. This work looks specifically towards training humans to perform a 2:3 polyrhythmic bimanual ratio using haptic force feedback devices (SensAble Phantom OMNI). We implemented an interactive training session to help participants learn to decouple their hand motions quickly. Three subjects (2 Females, 1 Male) were tested and have successfully increased their scores after adaptive training durations of under five minutes.

## I. Introduction

It has been widely accepted by the scientific community that performing two different, uncoordinated, tasks at the same time is challenging and even "almost impossible" for some tasks [1, 2]. Humans find it difficult to develop two unrelated thoughts simultaneously or carry out two cognitive tasks at once [3]. The same problem persists while attempting to do two physical distinct motions concurrently, such as: drawing a circle and a triangle at the same time [4] or playing with an "etch-a-sketch" [5].

On the other hand, motions which are dependent on each other tend to be easily performed by humans every day. These tasks/motions do not have to be symmetric, but they have to be part of a bigger sequence resulting in one distinct cognitive task. For example, peeling an orange (or opening a jar) requires two asymmetric tasks (one for each hand) resulting in a bimanual motion; however, both tasks are mutually dependent on each other.

For a human to easily perform two complex bimanual tasks, the tasks have to be tied to a common time base, which results in one sequence performed by two hands instead of two distinct sequences, one for each hand. In other words, both sequences have to be integrated to work together towards one goal, which results in one big sequence containing both subsequences [4]. For example, musicians (e.g., drummers, pianists) can perform two distinct sequences at the same time using both hands because both sequences are integrated into one time-based sequence played by both hands.

Research shows that the ability to master two different rhythms independently, does not infer the ability to simultaneously perform the rhythms using both hands [2]. Furthermore, the difficulty of performing the bimanual tasks cannot be fully perceived when doing a simple frequency rhythm (e.g., 1:1, 1:2, 1:3, etc.); however, it is perceived in nonharmonic frequency ratios (e.g., 2:3, 3:4, 5:7, etc.). These multi-frequency ratios are also called polyrhythms and are defined as when the terms are not integer multiples of each other.

Many models were created to explain this phenomena of having difficulty in performing polyrhythmic bimanual motions. Cattaert et. al., created a neural cross-talk computer model to proposed that neural cross-talk in motor execution results in the perceived difficulty [5]. Other research proposed that this difficulty is due to perceptual limitation, constraints on the regulation of timing or spatial position, symmetry bias towards spatial symmetry [1, 6, 7, 8].

This work focuses on building a haptic-based training system to enable humans achieve "decoupling" of their limbs and gain the ability to do polyrhythmic (complex frequency ratios) bimanual motions with their hands. Users are trained to perform complex polyrhythms using two SensAble Phantom OMNI haptic devices (fig. 1).

## II. Background

Other research has investigated the human ability to perform polyrhythmic bimanual coordination using haptic and visual feedback; however, almost all of the research done was focused on achieving multi-frequency ratios while using the feedback. In other words, the feedback was used during the evaluation period.

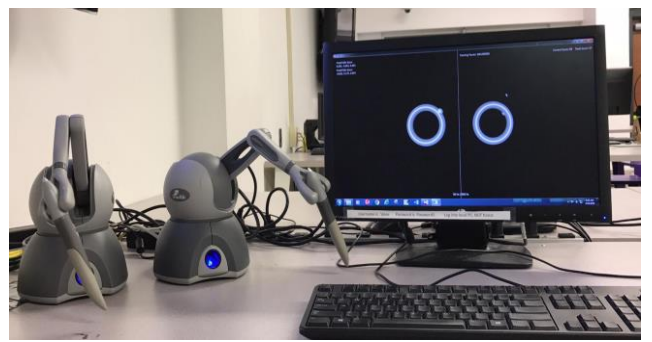

**Figure 1: Haptic based training system utilizing two SensAble Phantom OMNI devices**

It is important to note that this research focuses on using the feedback only in the training sessions to correct the users' hands trajectories and then evaluating the change of performance due to training and not while training.

### A. Haptic tracking

Research done by Rosenbaum et. al, looked specifically towards using bimanual haptic tracking to give users the ability to move both their hands independently from each other [3]. The authors designed three different experiments utilizing haptic tracking for bimanual coordination.

In the first experiment (fig. 2), participants were instructed to press two fingers, in each hand, against two buttons and stay in contact with the buttons as they move. The buttons are moved, by other humans (drivers), randomly to create unexpected, rapid bimanual motions. Participants were able to follow the bimanual motion of the buttons while blind folded.

• The authors are with the Department of Mechanical Engineering, University of South Florida, 4202 E. Fowler Ave., ENB118, Tampa, FL 33620. E-mail: ramy@mail.usf.edu, kylereed@usf.edu.

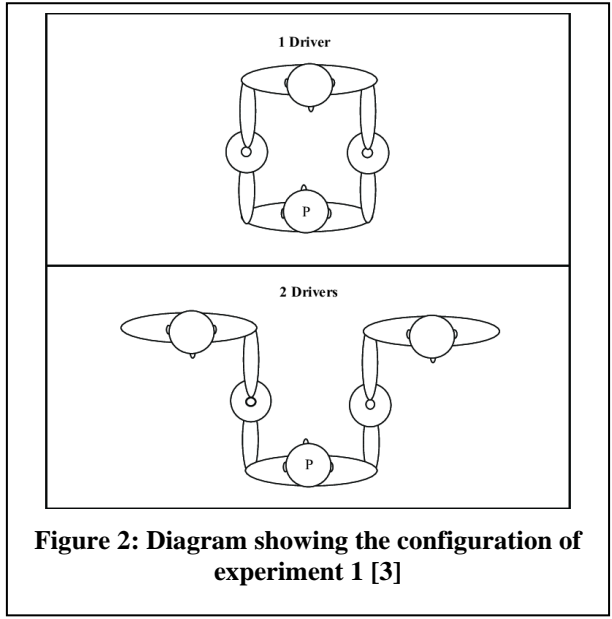

**Figure 2: Diagram showing the configuration of experiment 1 [3]**

In the second experiment, the authors used magnet for tracking on both sides of a vertical glass sheet. The magnet's force is insufficient to pull the participants' hands, but it was merely sufficient for guidance. Participants were able to move their hands independently to draw circles and squares concurrently.

In the third experiment, the authors instructed participants to do circular ratios of 1:1, 3:4 and 4:3 to avoid the strategic attention allocation to critical points in experiment 2 (being aware of the circular and square motions they were tracking). Participants were able to move their hand independently using haptic tracking.

It is worth noting that, in all of the three experiments, no visual feedback was provided to the participants; only haptic.

*B. Visual feedback*

Other research has explored the effect of providing visual feedback to participants while attempting to do multi-frequency bimanual coordination. The result is evaluated while the visual feedback is provided. In this case, feedback is not used for training; however, it is being used as a guidance system to explore the possibility of bimanual motion independence.

Kovacs et. Al., have tested on participants while providing a visual feedback in the form of Lissajous display along with a cursor displaying the position of their hands and a template providing information on the desired motion trajectory [9]. Participants were able to tune in a 5:3 bimanual coordination pattern and transfer to a 4:3 ratio with no previous practice.

Other research done by Boyles et. al., used visual feedback in the form on on-line relative velocity to help participants correct their relative motion in real-time [10]. In this experiment, the evaluation is based on the relative velocity instead of position.

Results from this method showed remarkably low coordination error, variability and biases. This method proved to be successful in guiding the participants to the correct bimanual motion for simple harmonic and multi-frequency bimanual ratios (1:2, 2:3, 3:4 and 4:5).

*C. Bimanual Rehabilitation*

Another direction being explored by researchers is the use of bimanual interactions for rehabilitation. In the work by McAmis and Reed, the authors studied bimanual tracking as the subjects follow a prerecorded trajectory with one arm and regenerate with the other arm for physical rehabilitation purposes [11, 12, 13].

This method is also being used for teaching physical skills because guiding the hand can prevent from learning due to the risk of performing the task passively. The research explored three different reference frames and ended up with the conclusion that visual symmetry and joint-space symmetry are easier than point mirror symmetry.

### III. METHODOLOGY

*A. Hardware Setup*

The hardware setup (fig. 1) for this experiment includes two SensAble Phantom OMNI devices and a computer with the Chai3D library installed. The OMNI devices have a force feedback workspace of 160 W x 120 H x 70 D mm with a footprint of 168 W x 203 D mm. This device can sense positions and apply forces in the x, y, z direction. The maximum exertable force at orthogonal arms position is 3.3 N and the stiffness ranges between 1.02 and 2.31 N/mm depending on the axis.

We used the Chai3D library to easily connect to the OMNI devices and control the forces applied by them. Chai3D is also convenient for designing the Graphical User Interface (GUI) to help participants visualize the task and locate the position of the end effector in real-time. The GUI does not give any visual feedback of the score to the subjects while in the testing mode. Also, no visual feedback of the desired trajectory is given to the users, only haptic feedback.

*B. Training Methods*

The goal of this experiment is to train participants on polyrhythmic ratios in order to achieve bimanual independence. Four training methods (table 1) were implemented to help users achieve better results in a short amount of time.

**Table 1: The four training methods**

| ID | Method | Non-Dominant Hand Training Power | Dominant Hand Training Power |
|---|---|---|---|
| 1 | Full | 100 | 100 |
| 2 | adaptive | 100 | 100-Current score |
| 3 | half | 100 | 0 |
| 4 | Zero (Test) | 0 | 0 |

The first training method (ID:1) is designed to force the participants to follow the training trajectory. In this method, the OMNI devices are applying full force (100) to move the end effector in circular motion, at a fixed velocity, and achieve the desired polyrhythmic bimanual ratio.

The second method (ID:2) forces the non-dominant hand to follow the desired trajectory with full power, while adaptively trains the dominant hand to perform the desired polyrhythmic ratio. The adaptive training method allows for the user and the trainer (OMNI) to work together; the power is calculated in real-time based on the current score. When the user's current score goes too low, the training power increases to force the dominant hand back on the right trajectory. This method allows participants to actively control the motion, make mistakes and learn from them as opposed to full control (ID:1), where the hand is usually passively guided by the feedback system.

The third method (ID:3) provides zero feedback to the dominant hand, while still providing full force feedback to the non-dominant hand. This method is used to illustrate to participants how the final testing session feels; it is not implemented to train the users and will not be used to calculate the final scores.

The final method (ID:4) is designed for testing; zero feedback is provided to the user. The total score is only calculated when this method is activated.

### C. Interactive Training Sessions

The actual training occurs in a session, where the training and testing pattern interactively adapts to the specific learning rate of each user. The Interactive Training Session (ITS) starts with 10 seconds of the adaptive training method (ID:2) and switches to testing mode (ID:4), where the total score is measured and recorded. After 20 seconds in testing mode, the score is evaluated. If the score is below 80% of the highest score, mode is switched back to 10 more seconds of adaptive training, otherwise, the user stays in testing mode.

This method is designed to dynamically change training and testing duration ratios based on the average score of the user in testing mode. Users who are learning and maintaining their score are expected to stay in the testing mode for longer periods of time. While users who are not learning the desired bimanual motion will keep alternating between testing and adaptive training indefinitely.

This method also helps users get back on the right trajectory in the event of losing the synchronicity between left and right hands; this method can indirectly function as a position reset.

### D. Scores Metric

The metric to calculate the scores can be divided in to several sections; relative position, relative velocity and total score calculation.

To calculate the relative position, an evaluation of the angle from the positive y-axes to the end effectors (clockwise) is necessary. It is important to note that for polyrhythmic ratios the angle calculation cannot be reset every 360 degrees, unlike rhythmic ratios, where this requirement is not needed for an accurate calculation.

For instance, for a ratio 2:3, if the non-dominant end effector is at 400 degrees, the dominant end effector needs to be at an angle of 600 degrees (240 degrees on the circle) for maximum score. Had the non-dominant end effector been reset to 40 degrees, the dominant would have been calculated to be at 60 degrees for the maximum score. A polyrhythmic ratio of 2:3 can reset its angle every 2 cycles (720 degrees) as opposed to a 1:2 to ratio (reset every 360 degrees).

After calculating the angles for the two end effectors, we compute the absolute value of the difference between them and consider the outcome to be the error. To convert the error to a 0-100 based score, we assumed the maximum error to be at an angle difference of 180. The final score calculation of the relative position – after simplifications – is:

$$Score = \left(1 - \frac{current\ angle - desired\ angle}{180}\right) * 100$$

The current angle is the angle evaluated at the dominant hand end effector, while the desired angle is calculated by evaluating the angle of the non-dominant hand end effector and multiplying by the ratio. The current score is calculated as the moving average of the scores over a window of 20 frames. While the total score – only calculated while testing – is the average of all frames from the beginning of the testing session.

The relative velocity is also used in the calculation of the total score. The OMNI provides information on the linear velocities. We use the square root of the sum of squares (2-D magnitude equation) on the Y and Z axes to calculate the magnitude of the velocity of each end effector. We used a moving average for each of the end effectors (window = 40), then divided the dominant's hand velocity by the non-dominant's hand velocity to find the actual relative velocity.

The error in relative velocity is computed as the current relative velocity minus the desired relative velocity (3/2 for a 2:3 ratio). To convert this error value to a 0-100 score we used the following mapping function:

$$Score = A * e^{-Bx^2}$$

Where A = 100, B=1. The variables A and B are based on maximum relative velocity error observed from the data.

The total score combining relative position and relative velocity is the average of both metrics as shown in the following equation:

$$Total\ Score = (R. Position * 0.5) + (R. Velocity * 0.5)$$

This metric doesn't yield a score that depends on radius of the circles or absolute velocity. The score only depends on the relative positions (angles) and the relative velocities.

### E. Experimental Setup

Three human subjects were tested in this experiment. The subjects were given enough time to familiarize themselves with the setup. All participants were instructed to continue to draw circles while in the testing mode. A polyrhythmic ratio of 2:3 was used with all of the test subjects. All of the participants experienced one interactive training session. Table 2 summarizes the participants' demographic information.

**Table 2: Demographic information for the three participants**

| ID | Gender | Age | Dominant Hand | Ratio |
|----|--------|-----|---------------|-------|
| 1  | F      | 46  | right         | 2:3   |
| 2  | F      | 27  | right         | 2:3   |
| 3  | M      | 51  | right         | 2:3   |

Results for each participant are presented in the following section.

## IV. RESULTS AND DISCUSSIONS

The results for this experiment include a visualization of the position, velocity and combined total score versus time for each participant. A percentage change chart is also provided to analyze the trend of increasing/decreasing of the total score.

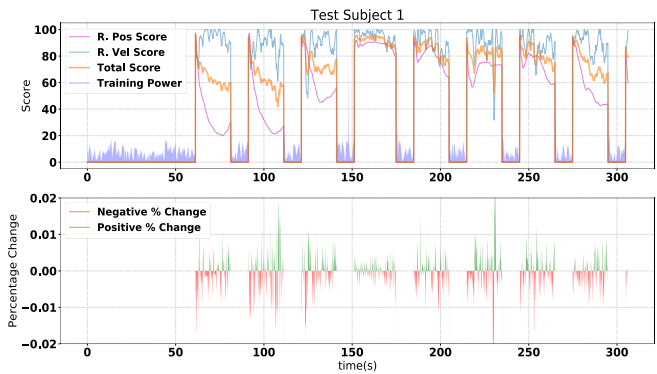

**Figure 3: Test Scores and percentage change for subject 1**

The results (fig. 3) for test subject 1 show an increase in the average score over time. The fourth testing session also shows high scores between 80 and 100 as well as the increase in duration of this specific session compared to the other testing session.

A closer look at the first testing session will reveal the importance of calculating the relative velocity. In this case, the relative position dropped down to a score of 20 while the relative velocity stayed in the 80-100 range. It is possible for the users to correct their scores in the middle of a testing session by only adjusting the relative velocity, which merely affects the relative velocity score. Calculating the relative velocity and averaging it with the relative position will provide a more accurate score.

**Table 3: Repeated measures ANOVA and POST-HOC statistical analysis results for subject 1**

|   | 2 | 3 | 4 | 5 | 6 | 7 | 8 |
|---|---|---|---|---|---|---|---|
| 1 | 0 | 0 | 0 | 0 | 0 | 0 | 0 |
| 2 |   | 0 | 0 | 0 | 0 | 0 | 0 |
| 3 |   |   | 0 | 0 | 0 | 0 | 0 |
| 4 |   |   |   | 0 | 0 | 0 | 0 |
| 5 |   |   |   |   | 0 | 0 | 0 |
| 6 |   |   |   |   |   | 0 | 0 |
| 7 |   |   |   |   |   |   | 0 |

$F(7,3486)=1794.381$, $P = 0.000$

Table 3 shows the statistical analysis for the eight testing sessions by subject 1. A Repeated Measures ANOVA was evaluated for all of the eight testing sessions and resulted in a P value of zero. A follow-up post-HOC test resulted in a pairwise comparison chart for the eight factors (Table 3). None of the factors showed a statistical difference with any other factor (<0.05).

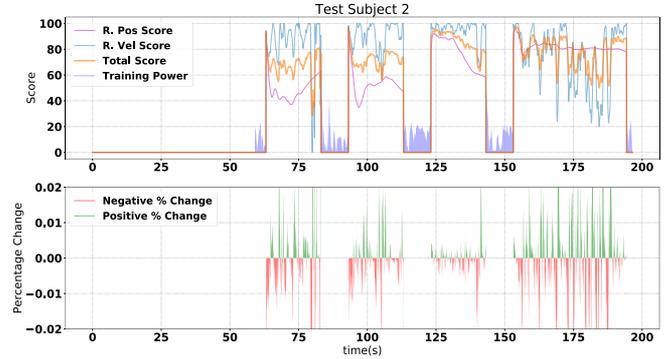

**Figure 4: Test Scores and percentage change for subject 2**

The results (fig. 3) for test subject 2 show a much better learning curve; the time intervals of testing significantly increased over time. She was able to maintain her score above the 80% threshold for longer periods of time. It is noticeable that more training power was used during her adaptive training sessions, than the training power used for subject 1.

**Table 4: Repeated measures ANOVA and POST-HOC statistical analysis results for subject 2**

|   | 2 | 3 | 4 |                              |
|---|---|---|---|------------------------------|
| 1 | 0 | 0 | 0 | $F(7,3486)=1794.381$, $P = 0.000$ |
| 2 |   | 0 | 0 |                              |
| 3 |   |   | 0 |                              |

The ANOVA and post-HOC statistical analyses show that the mean values of each of the testing sessions statistically differ from each other. The ANOVA resulted in a P value of zero, and a follow up post-HOC shows that none of the factors means have a P value greater than 0.05 with any other factor mean.

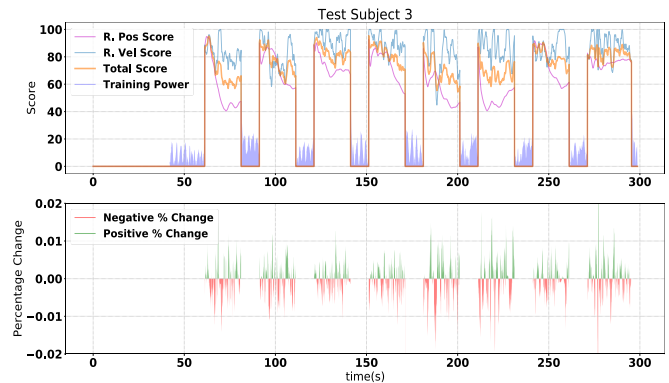

**Figure 5: Test Scores and percentage change for subject 3**

The results (fig. 5) for subject 3 show that the mean scores also slightly increased over time. The subject was able to learn to maintain his scores for longer periods of time. The last two testing sessions resulted in a significantly higher score than the first session.

**Table 5: Repeated measures ANOVA and POST-HOC statistical analysis results for subject 3**

|   | 2 | 3 | 4 | 5 | 6 | 7 | 8 |
|---|---|---|---|---|---|---|---|
| 1 | 0 | 0 | 0 | 0 | 0 | 0 | 0 |
| 2 |   | 0.02 | 0 | 0 | 0 | 0 | 0 |
| 3 |   |   | 0.073 | 0 | 0 | 0.779 | 0 |
| 4 |   |   |   | 0 | 0 | 0.007 | 0.001 |
| 5 | F(7,3479)=769.844, P = 0.000 |   |   |   | 0 | 0 | 0 |
| 6 |   |   |   |   |   | 0 | 0 |
| 7 |   |   |   |   |   |   | 0 |

The ANOVA test resulted in a P value of zero, so a follow up post-HOC was carried out. The results show no significant difference between the means of testing sessions three and four and between three and seven. All the other testing sessions are statistically different.

The statistical difference between the first and last testing sessions for all of our subjects is a good indicator that they are learning to maintain and increase their scores over time.

We evaluated the Pearson correlation between all of the subjects' total scores and found a correlation of 1.0 between test subjects one and three. We found a correlation of 0.722 between subjects one and two, and a correlation of 0.826 between subjects two and three. The correlation values support the evidence from the scores' charts that the test subject two has learned the most. Subject two is nearly half the age of subjects one and three which can influence the learning rate. This could be the reason why subject two learned much faster and only needed four training sessions, as compared to eight sessions for the other subjects.

The three subjects have experienced a sharp initial drop in the scores during the first testing session; however, they have successfully learned to significantly reduce the slope of this drop and maintain their scores for longer durations. This is another good indicator that there is a learning process and it is not impossible to learn polyrhythmic bimanual motions in relatively short amount of time.

Subject two reported that she attempted to learn visual cues from the training sessions to help her while testing, but she found concentrating on the relative velocity "much easier" and only used the relative velocity approach starting from the second testing session.

## V. CONCLUSION AND FUTURE WORK

In this experiment, we used force feedback to train three human subjects to perform a complex polyrhythmic bimanual motion. The three Interactive Training Sessions (training + testing) were under five minutes in duration; however, participants have learned to maintain and increase their scores.

We plan on doing more ITS scheduled twice a week for a period of two months and see the results. We also plan on recruiting more people to form a larger, balanced set of subjects of different ages. We are currently collecting data using tablet and smartphone devices to determine how well people can perform certain polyrhythm patterns [14].